\documentclass[aps, pra, superscriptaddress, twocolumn, longbibliography]{revtex4-1}
\usepackage[utf8]{inputenc}
\usepackage[margin=0.6in]{geometry}
\usepackage{graphicx}
\usepackage[table,xcdraw]{xcolor}
\usepackage{amsmath,amssymb,mathtools}
\usepackage{booktabs}
\usepackage{multirow}
\usepackage{braket}
\usepackage{blindtext}
\usepackage{float}
\usepackage{gensymb}
\usepackage{subcaption}
\usepackage[export]{adjustbox}

\usepackage[toc,page]{appendix}
\usepackage[colorlinks=true,pdfborder={0 0 0},linkcolor=blue, citecolor=blue, allcolors = blue]{hyperref}

\begin{document}

\renewcommand{\abstractname}{\vspace{-\baselineskip}}

\title{Endoscopic fiber-coupled diamond magnetometer for cancer surgery}

\author{A. J. Newman}
\email{Alex.Newman@warwick.ac.uk}
\author{S. M. Graham}
\author{C. J. Stephen}
\affiliation{\footnotesize Department of Physics, University of Warwick, Gibbet Hill Road, Coventry, CV4 7AL, United Kingdom}
\affiliation{\footnotesize EPSRC Centre for Doctoral Training in Diamond Science and Technology, University of Warwick, Coventry CV4 7AL, United Kingdom}

\author{A. M. Edmonds}
\author{M. L. Markham}
\affiliation{\footnotesize Element Six Innovation, Fermi Avenue, Harwell Oxford, Didcot OX11 0QR Oxfordshire, United Kingdom}

\author{G. W. Morley}
\email{Gavin.Morley@warwick.ac.uk}
\affiliation{\footnotesize Department of Physics, University of Warwick, Gibbet Hill Road, Coventry, CV4 7AL, United Kingdom}
\affiliation{\footnotesize EPSRC Centre for Doctoral Training in Diamond Science and Technology, University of Warwick, Coventry CV4 7AL, United Kingdom}

\begin{abstract}

Interoperative measurements using magnetic sensors is a valuable technique in cancer surgery for finding magnetic tracers. Here we present a fiber-coupled nitrogen-vacancy (N-V) center magnetometer capable of detecting iron oxide suspension (MagTrace\texttrademark{} from Endomagnetics Ltd.) used in breast cancer surgeries. Detection of an iron mass as low as 0.56~mg has been demonstrated, 100 times less than that of a recommended dose at a maximum distance of 5.8~mm. Detection of an iron concentration as low as 2.8 mg/ml has also been demonstrated, 20 times less than a recommended dose. The maximum working distance from the sensor can be as large as 14.6~mm for higher concentrations. The sensor head has a maximum diameter of 10~mm which would allow it to be used for endoscopy, laparoscopy and interoperative surgery. 

\end{abstract}

\maketitle

\section{Introduction} 

Non-toxic and non-radioactive techniques for finding metastasized breast cancer in lymph nodes are becoming more prevalent in hospitals around the world as magnetic based sensors have developed and are able to be used in place of blue dyes and radioisotopes \cite{MagneticSensorSLNBsuperparamagneticBeads_Kitamoto2012, SLNBSupermagneticironoxideJapan_Taruno2018, ApplicationMagnetcNanoParticlesSLNBReview_Liu2022, magneticSLNBExample1_Zada2016, magenticSLNBExample2_Anninga2016, magenticSLNBExample3_Douek2014}. Radioisotopes, detected using a gamma probe, are expensive and not available in all hospitals. Furthermore, using radioisotopes exposes the medical team and the patient to radiation which is preferably avoided. Blue dye can be injected into the primary tumor and accumulates in the sentinel lymph node over time, highlighting its location to the surgeon through visual inspection. Allergic reactions to blue dye have been reported in approximately one in a hundred patients \cite{BlueDyeAlergicReactionStudy_Bezu2011, BlueDyeAlergicReactionStudy2_Thervarajah2011, BlueDyeAlergicReactionStudy3_Albo2011, BlueDyeAlergicReactionStudy4_Cimmino2001, BlueDyeAlergicReactionStudy5_Miklos2001}. Blue dyes also can cause long term staining of the skin at the injection site.

To mitigate the risks associated with these techniques, magnetic sensors have been increasingly utilized in sentinel lymph node biopsy (SLNB) to provide detection of non-toxic superparamagnetic iron oxide nanoparticles (SPION) \cite{LaparaoscopicProbeMagneticNanoParticles_Loosedrecht2021, NICE_sentimagRecomendation, developmentMagneticParticleSLNB_Inagaki2021, MagneticParticleImagingForSLNB} suspended in liquid which can be injected into the patient from 20 minutes to two weeks before surgery. Magnetic detection can have two operating modes, one using alternating current (AC) and a coil to generate an AC magnetic field \cite{magneticNanoparticlesImaging_ACCoil_Grafe, magneticNanoparticlesImagingACCoils_Kuwahata2017, magneticNanoparticlesImaging_ACCoilSingleSide_Sattel2012} and the other using permanent magnets \cite{magneticNanoparticlesImaging3DPermMagnetHallProbe_Kuwahata2017}. AC field methods cause the target to produce an AC magnetic field which is detected. Heating of tissue due to the frequency and amplitude of the nearby probe through Joule heating is also an issue. \cite{HandheldMagneticProbePermMagnetSLNB_Sekino2018, ACMagneticProbe_JouleHeating2024, ACMagneticProbe_JouleHeating2014}.

NV center magnetometry couples high dynamic range \cite{TensorGradiometryDiamondMagnetometer_Alex2024, cleversonVectorDynamicRange, KumarDynamicRange} and high magnetic sensitivity \cite{Barry2020SensitivityMagnetometryb, Taylor2008High-sensitivityResolution, stuarts_sensitivity, Zhang_sensitivity} with small sensing volumes e.g. $\mathrm{1~mm^3}$ or less \cite{TensorGradiometryDiamondMagnetometer_Alex2024, stuarts_sensitivity, Acosta2009DiamondsApplications, Toyli2013FluorescenceDiamond, Doherty2013TheDiamond, Hsieh2019ImagingSensor}. The small sensor size is appropriate for SLNB applications because it reduces the volume inserted into the patient reducing trauma and bruising. Furthermore for laparoscopic surgeries, probes are limited in size by the trocar used in these procedures, typically to 12 mm diameter, which highlights the need for small probes. \cite{ExperimentalStudyTrocarLaparoscopic_Shiran2020}. The ability to fiber couple the diamond also provides more maneuverability when compared with a solid rod or rigid baton found in other devices \cite{Kuwahata2020MagnetometerApplications, HandheldMagneticProbePermMagnetSLNB_Sekino2018}. Sensing of magnetic fields using NV centers in diamond has previously been used to detect magnetic nanoparticles for sensing in biomedical tissue \cite{Kuwahata2020MagnetometerApplications, Davis2018MappingMagnetometry}. One design had a non-mobile sensor head for imaging cells that had been removed from the patient or animal \cite{Davis2018MappingMagnetometry}. The other had a sensor head that is 20~mm in diameter which is too large for these surgical applications, as well as an AC sensitivity of 57~$\mathrm{nT/\sqrt{Hz}}$~\cite{Kuwahata2020MagnetometerApplications}.  

\begin{figure}[t]
    \begin{subfigure}[t]{0.47\textwidth}
        \phantomsubcaption
        \label{fig:setupDiagram} 
        \includegraphics[width = \textwidth]{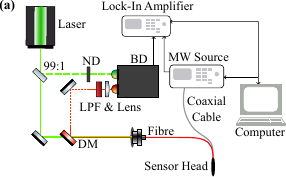}
    \end{subfigure}%

    \vspace{0.25cm}
    \begin{subfigure}[t]{0.47\textwidth}
    \phantomsubcaption
    \label{fig:sensorDiagram} 
    \includegraphics[width =\textwidth]{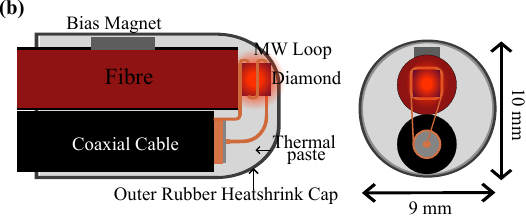}
    \end{subfigure}

    \begin{subfigure}[t]{0.45\textwidth}
    \phantomsubcaption
    \label{fig:probde_head_picture} 
    \includegraphics[width =\textwidth]{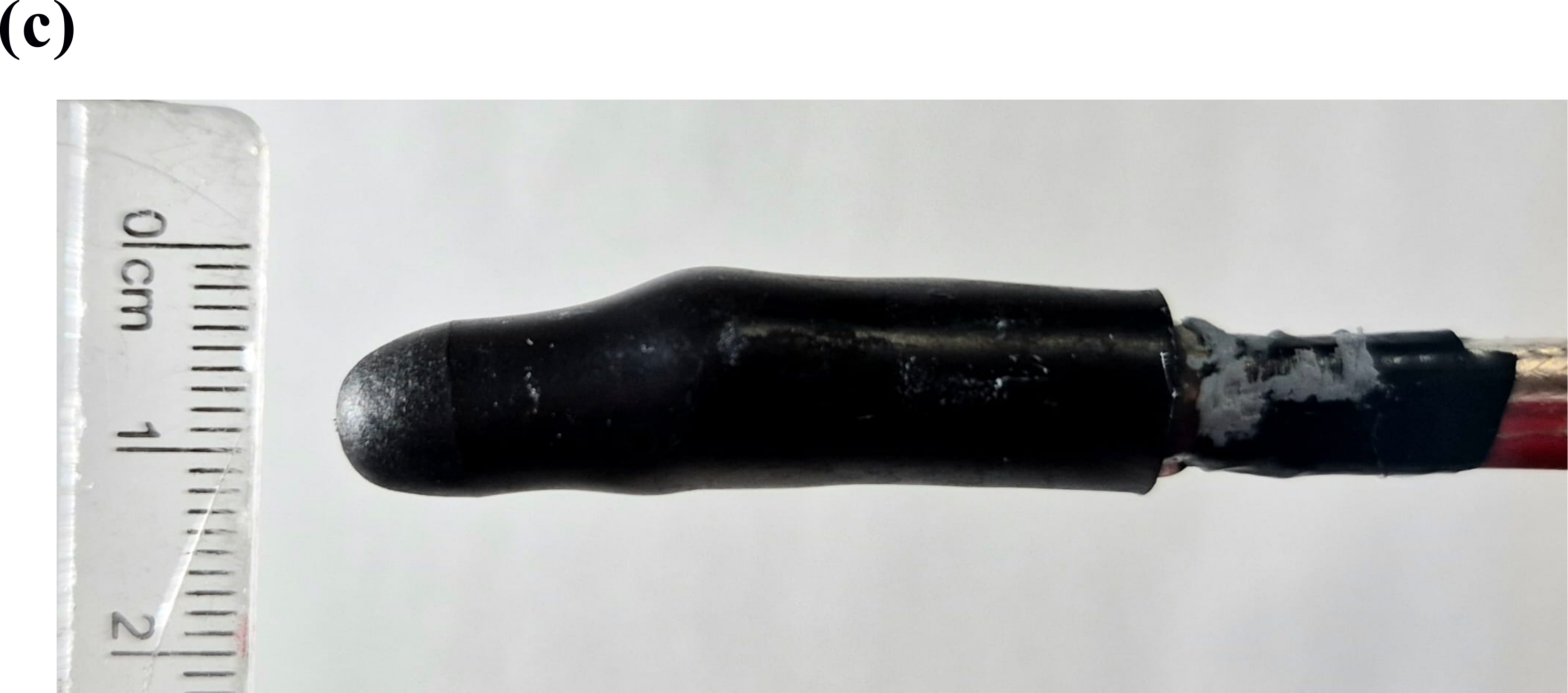}
    \end{subfigure}
    \caption{(a) Diagram of the scanning magnetometry setup (ND - neutral density filter, BD - balanced detector, DM - dichroic mirror,  MW - microwave, LPF - long-pass filter). (b) The diamond is directly coupled with glue to the end of a polished fiber. An MW loop is attached and glued in place and soldered to a coaxial cable. A bias magnet is placed and secured in place to provide resonance splitting. (c) Photo of the probe head with ruler for scale. }
\end{figure} 

Here we present an endoscopic fiber-coupled diamond magnetometer with a maximum diameter of 10~mm. This diameter makes the probe applicable for use in endoscopic, laparoscopic, and interoperative surgical procedures. Its compact design is achieved by using a small permanent magnet, that is attached to the probe head, for both a bias field and inducing fields in the sample. This eliminates the need for bulky AC excitation and cancellation coils, as well as the need for providing power to the sensor head. The probe can detect iron oxide nanoparticles at concentrations as low as 2.8 mg/ml at a distance of 5.8~mm, with a maximum detection range of 14.6 mm at the highest tested concentration of 28 mg/ml. The system has an unshielded sensitivity of $\mathrm{12.3 \pm 4.1~nT/\sqrt{Hz}}$ between 0.5 and 10~Hz. Unlike conventional NV magnetometers that require a stationary permanent magnet bias field, our design incorporates a bias magnet attached to the probe, allowing it to move without losing sensitivity. This feature is essential for handheld magnetometer probes in applications such as medical surgery.



\section{Experimental Detail and Method}

The optoelectronic setup is shown in Fig. \ref{fig:setupDiagram}. A Laser Quantum 532-nm GEM laser is used for excitation of the NV ensemble in the diamond. A laser power of 250~mW is used with approximately 150~mW measured at the diamond, after accounting for losses through the system including the beam splitter and fiber-coupling efficiency. The optical system is inside a rack on wheels for increased mobility. The fiber used is a Thorlabs FG910UEC. The fiber is 1.5~m long with a core diameter of 910~$\mathrm{\mu m}$ (0.22~NA) and with a steel FCPC (ferrule-connector-physical-contac) connector on one end and a cut bare fiber exposed on the other end. The diameter of the fibre is larger than the size of the diamond to increase light collection. The FCPC connector is fastened to a Thorlabs SM1FC 2.2~mm wide key FCPC terminated fiber adapter inside the optics box. The fiber is fed through and out the back of the optics box. The diamond is directly coupled to the bare-fiber end, seated and secured in place using glue (Loctite 401). During the seating process, the laser was on and photoluminescence (PL) was collected from the diamond. While the glue was still viscous, the diamond was gently adjusted using tweezers to maximize the PL collection, measured using a PicoScope 5442D Series oscilloscope. Once the best PL level was achieved, the laser was turned off and the glue was left to cure. A coaxial cable was attached to the side of the fiber, running the length of the fiber but separating off before the fiber goes into the optics box, with the coaxial cable connecting to the microwave source. The coaxial cable has a SMA (SubMiniature version A) adapter on one end and exposed metal shielding and copper core on the other. The coaxial cable was attached to the fiber such that the exposed copper core of the coaxial cable was as close as possible to the diamond. A length of 0.4~mm diameter copper wire was wrapped fully around the diamond with one turn, and the ends were soldered to the metallic shielding and the core of the coaxial cable. Microwaves were then sent to the probe head through the copper wire to produce an ODMR spectrum, measured by the PicoScope using the LIA scaled demodulated output. The copper wire was adjusted in shape using tweezers to maximize the ODMR contrast. Further glue was then used to secure the copper wire in place around the diamond. The tip of the fiber, including the diamond and coaxial cable is coated in thermal paste to help with heat transfer away from the diamond to prevent the glue from failing and to increase PL collection by using the paste to scatter light into the fiber. The fiber end and coaxial cable is then covered with a heatshrink cap (HellermannTyton Polyolefin Cross Linked (POX) 10~mm expanded diameter) to protect the diamond, copper loop and thermal paste as well as make the device light-tight. The resulting maximum diameter is 10~mm.

\begin{figure}[t]
    \centering
    \includegraphics[width=\linewidth]{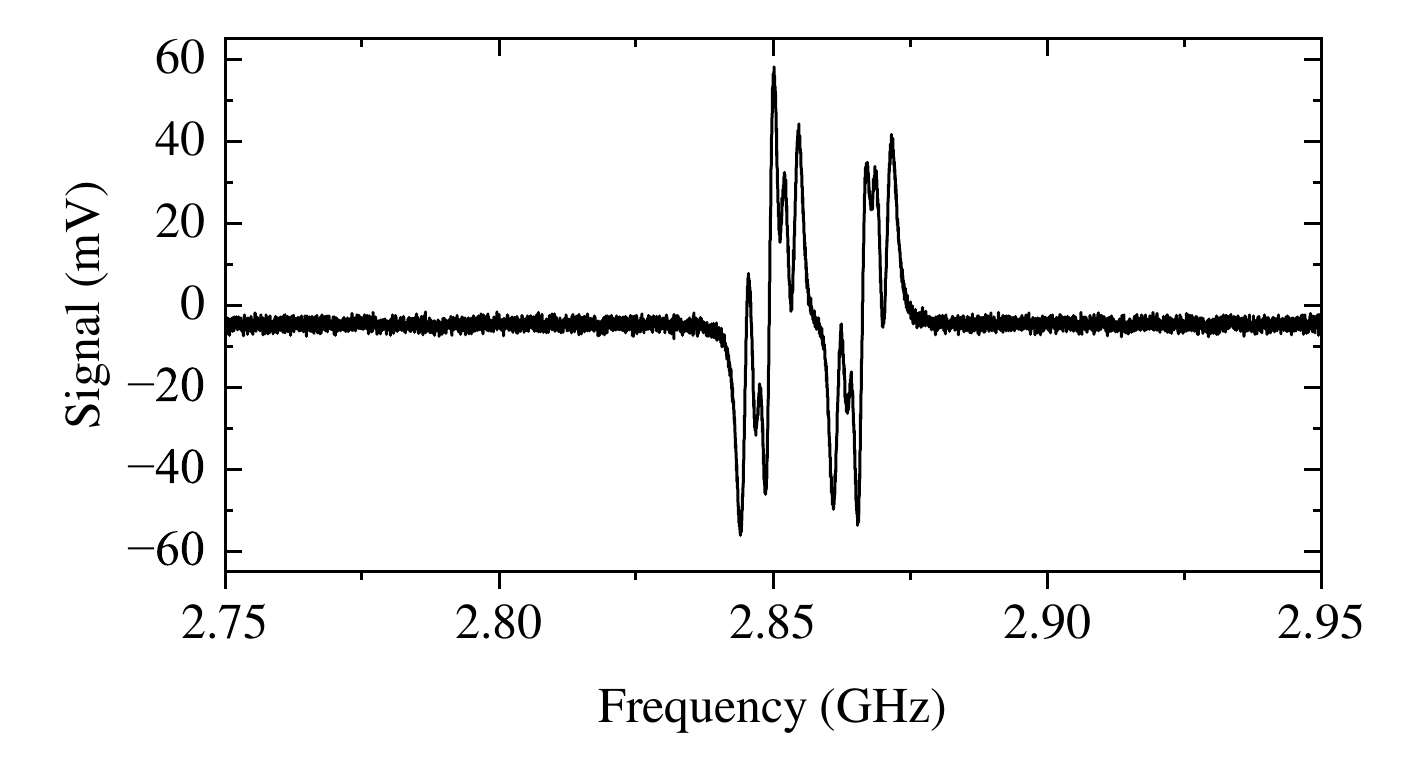}
    \caption{Optically detected magnetic resonance spectrum after the neodymium magnet was aligned close to a [100] orientation and secured in place.}
    \label{fig:ODMR}
\end{figure}

\begin{figure}[t]
    \centering
    \begin{subfigure}[t]{0.5\textwidth}
        \phantomsubcaption
        \label{fig:stiched}
        \includegraphics[width = \textwidth]{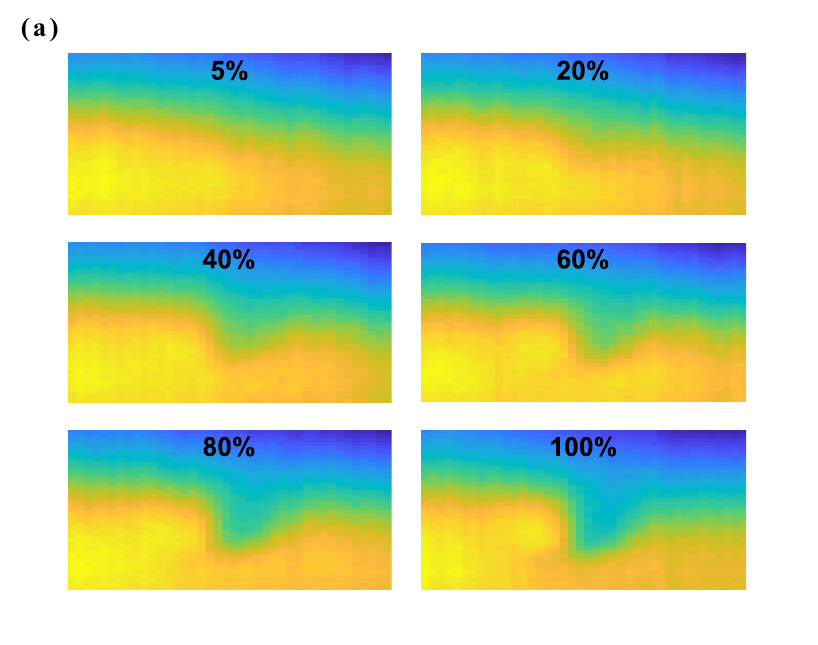}
    \end{subfigure}
    \begin{subfigure}[t]{0.44\textwidth}
        \phantomsubcaption
        \label{fig:LineProfile}
        \centering
        \includegraphics[width = 1\textwidth]{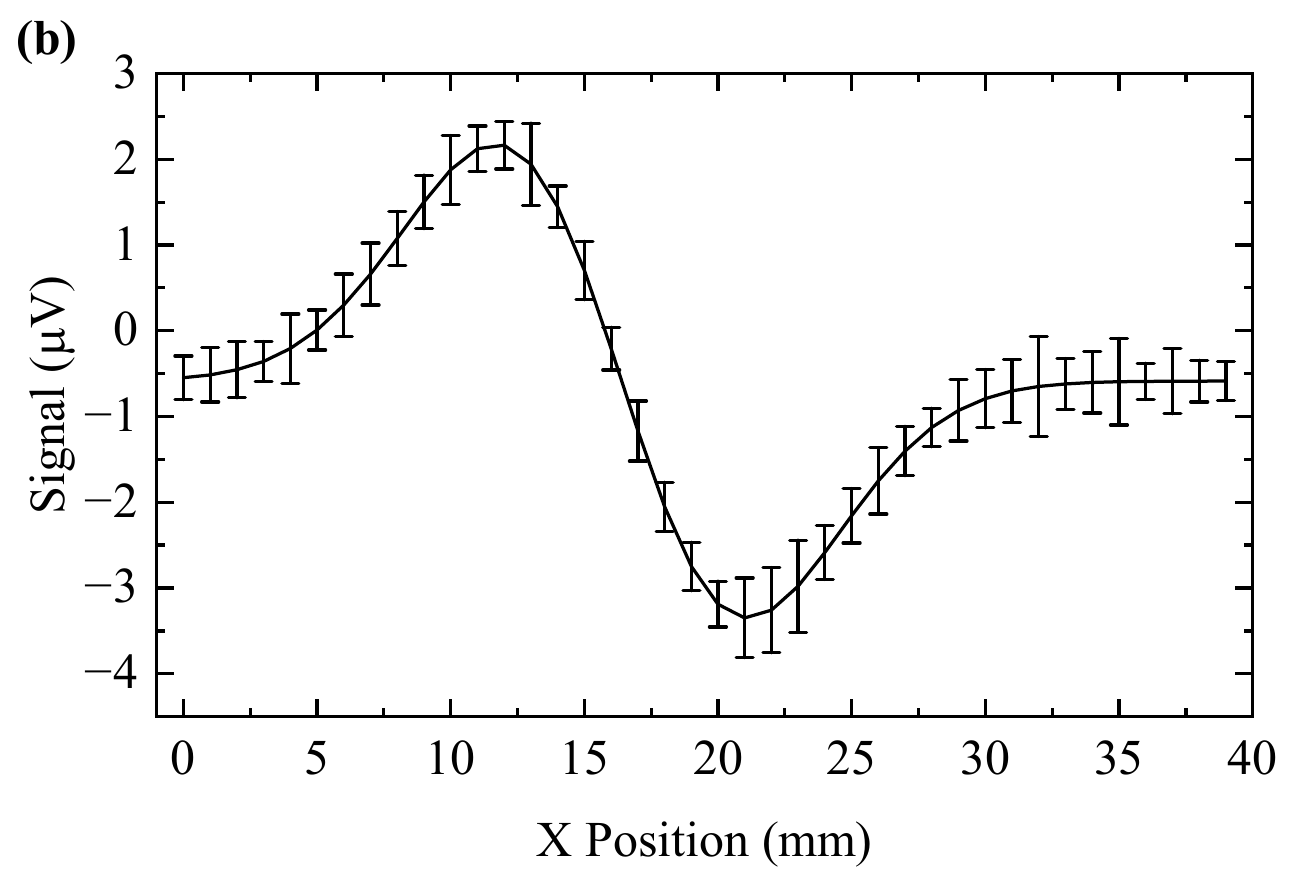}
    \end{subfigure}
    \begin{subfigure}[t]{0.44\textwidth}
        \phantomsubcaption
        \label{fig:VerticalContrast}
        \centering
        \includegraphics[width = 1\textwidth]{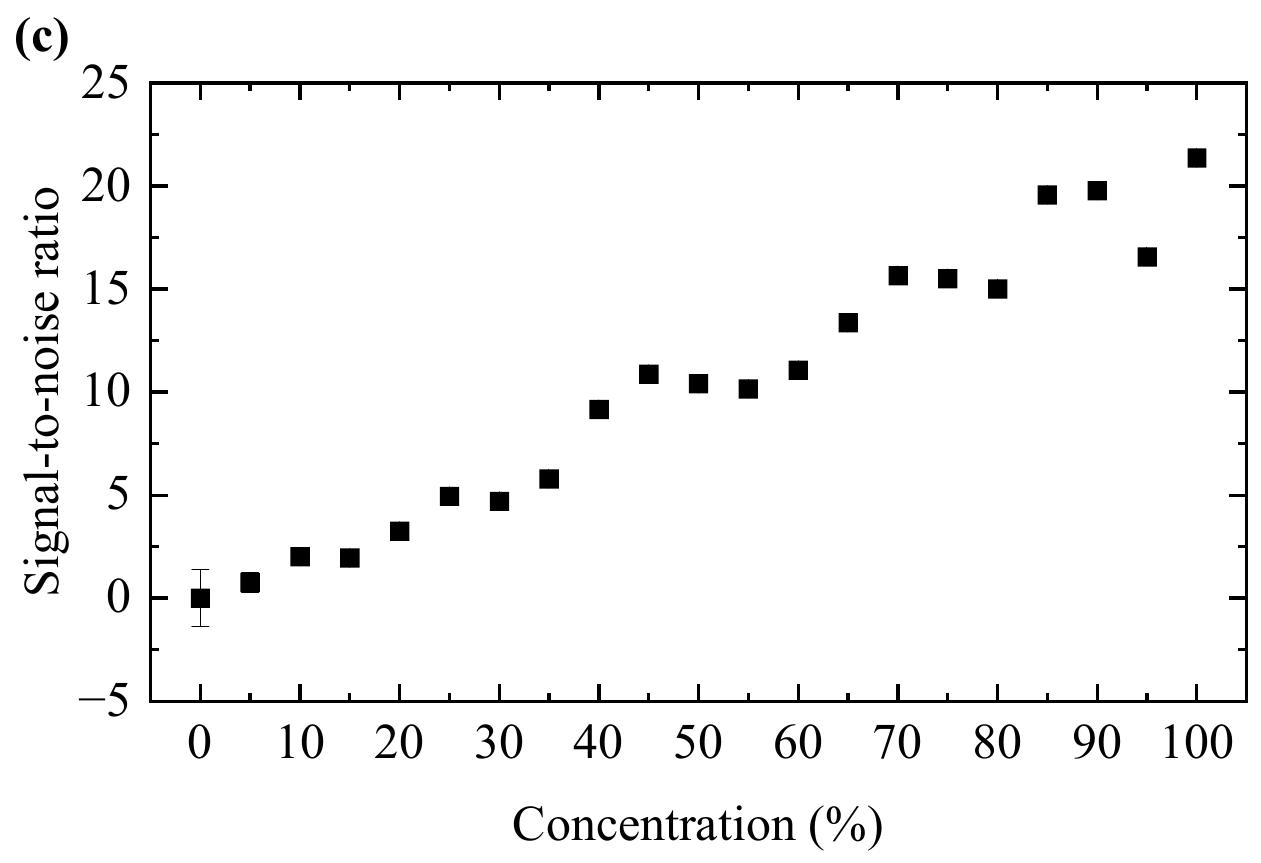}
    \end{subfigure}
    \caption{Measurement contrast decreases with concentration. (a) 40 $\times$ 40 mm magnetic image scans for a selection of concentrations. Each scan took 8 minutes. (b) X line profile for 80\% concentration. (c) Contrast of each concentration using the maximum and minimum of the Gaussian derivative fits.}
\end{figure}

Both the laser excitation light and red PL from the diamond travel through the same fiber. The PL from the diamond is separated out using a dichroic mirror (Thorlabs DMSP650) to be focused using an aspheric lens, onto one of the two photodiodes of a balanced detector (Thorlabs PD450A). The other photodiode takes a reference beam from the laser, directed using a beam sampler and mirror and passing through a neutral density (ND) filter. This is done to cancel common mode noise from the laser. The input power of both photodiodes is balanced when the microwaves are on resonance to get the best cancellation when magnetically sensitive. The balanced detector output is passed to a Zurich MFLI DC 500-kHz lock-in amplifier (LIA). The microwaves are provided by an Agilent N5172B. A microwave power of 10~dBm was used for all data taken in this paper. The diamond used was a 0.5-mm-cube, low-strain, 99.995\%-$^{12}\mathrm{C}$ chemical-vapor-deposition (CVD)-grown diamond by Element Six, polished on all sides. 

A single N42 neodymium-iron-boron 2.5 $\times$ 7 $\times$ 2.5~mm magnet is attached to the side of the end of the fiber. An optically detected magnetic resonance spectrum (ODMR) is used to determine where best to place the magnet to provide close to a [100] alignment, so the four peaks for the $\mathrm{m_s} = 0$ to $\mathrm{m_s} = -1$ transition overlap and the four peaks for the $\mathrm{m_s} = 0$ to $\mathrm{m_s} = +1$ transition overlap (Fig. \ref{fig:ODMR}). A [100] alignment was chosen to provide approximately a 2.3 times improvement in sensitivity along the [100] axis, considering a 4 times increase in contrast but also considering the projection of the magnetic field along the NV symmetry axes giving a 0.58 times decrease in measured field. The magnet was then secured in place using glue. 

The completed sensor setup was attached to a 3D printing stage to scan the sensor head in a controlled XY plane and to accurately adjust the height of sample scans. A step size of 1.0~mm was used for the X and Y axis movement.



\section{Results and Conclusions}

The original MagTrace\texttrademark{} (Endomagnetics Ltd) suspension comes in a 2~ml vial containing approximately 55~$\pm$4~mg of iron in the form of iron oxide and 64~mg of carboxydextran. These are combined to form carboxydextran-coated superparamagnetic iron oxide. This gives an approximate concentration of 28~mg/ml of iron, in the form of maghemite (the gamma phase of iron oxide $\mathrm{\gamma-Fe_2O_3}$) in the suspension. By measuring out specific volumes of this suspension and mixing with water to dilute it, varying concentrations of this suspension were made to test the sensitivity of the magnetometer to changes in iron oxide concentration.

To determine the minimum concentration of iron oxide suspension the sensor could detect and at what distance, twenty 200~$\mathrm{\mu}$l vials were filled with increasing concentrations of iron oxide suspension. The concentrations started at 0\% and increased by 5\% up to 100\%. The concentrations were made using a mix of MagTrace\texttrademark{} (Endomagnetics Ltd) suspension and water, measured out using an electronic pipette. A 3D printed jig was made to hold the samples, with a hole the same diameter as the vials in the centre, secured onto the 3D printing stage. This allowed the swapping out of vials while keeping their position in the scan constant. The sensor head was scanned over the same area each time to keep measurements consistent. The scanned area was 40~$\times$~40~mm with a separation distance of 0.5~mm between the top of the vials and the sensor tip. The magnetic image for each concentration shows the change in lock-in amplifier signal output, as a function of position, caused by the proximity of the magnetic material shifting the magnetic resonance frequency of the NV centers in the diamond (Fig. \ref{fig:stiched}). For each point in the scan, a measurement was made with a 0.5~s acquisition time at a sampling rate of 13.39~kSa/s. The standard deviation of each acquisition was also recorded and used as the error of each pixel.



To confirm the detection of the varied concentrations, line profiles were taken both along the X and Y axis, across the spots seen in the image (Fig. \ref{fig:LineProfile}). Due to the dipole shape of the images, only the horizontal line profile was used to calculated the signal-to-noise ratio (SNR) of the measurement. This was because the vertical line profile passed through the local minimum of the dipole and provided no signal change. 

The line profiles showed a derivative Gaussian profile and a derivative Gaussian function was used to fit the data. The SNR was calculated for the individual line profiles by using the maximum value of the curve fit minus the minimum value of the curve fit, divided by the signal standard deviation (Fig. \ref{fig:VerticalContrast}). The lowest concentration with a SNR above 1, was 10\% concentration, with an SNR of 2.0 $\pm$ 0.3. This was taken as the lowest concentration that could be reliably detected. 10\% concentration of the 200~$\mathrm{\mu}$l vial used equates to a volume of 20~$\mathrm{\mu}$l. This equates to an approximate iron mass of 0.56~mg. Therefore, the sensor has shown to be capable of detecting 0.56~mg of iron in 200~$\mathrm{\mu}$l of suspension or 2.8 mg/ml, at a distance of 0.5~mm from the sample. 


\begin{figure}[t]
    \centering
    \begin{subfigure}[t]{0.46\textwidth}
        \phantomsubcaption
        \label{fig:SignalVsDistance}
        \includegraphics[width = \textwidth]{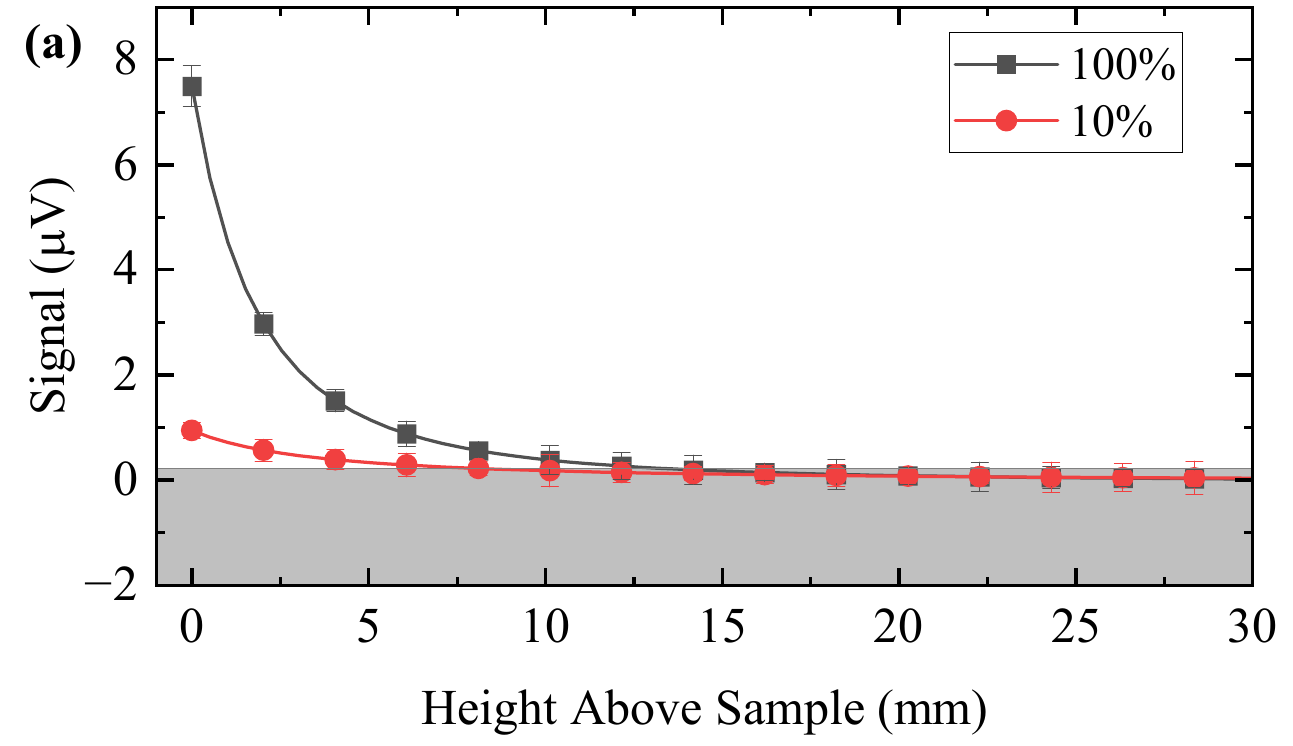}
    \end{subfigure}
    \begin{subfigure}[t]{0.46\textwidth}
        \phantomsubcaption
        \label{fig:MaxDistance}
        \centering
        \includegraphics[width = 1\textwidth]{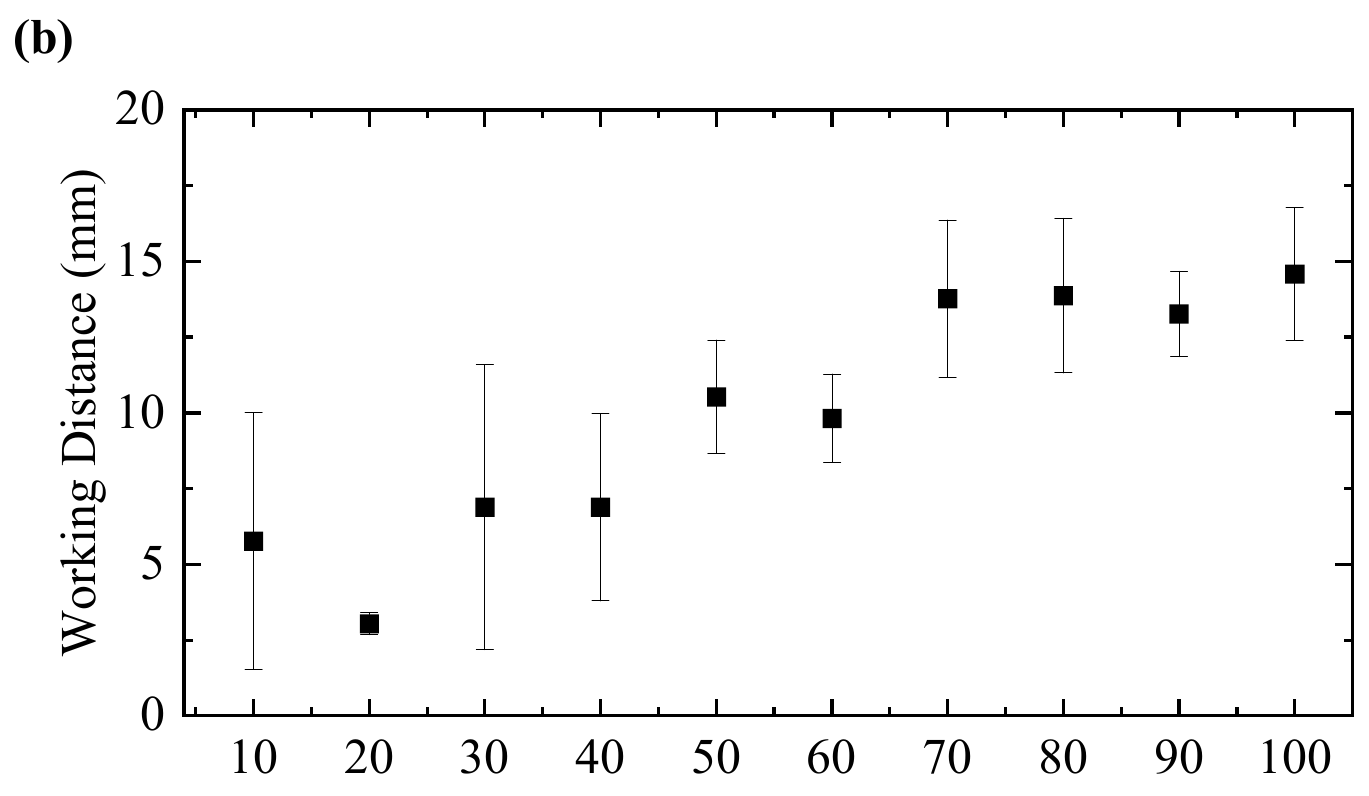}
    \end{subfigure}
    \caption{The rate at which the signal changes with distance varies with iron oxide concentration in the suspension. (a) The change in lock-in amplifier (LIA) output voltage as the sensor head is moved away from the sample. A power-law fitted curve is shown for 100\% (black) and 10\% (red). The gray area shows the noise floor. (b) The maximum working distance, the point at which the fitted power-law curve increased by one standard deviation, for each concentration. }
\end{figure}

To determine the maximum separation distance at which each concentration could be detected, each concentration was placed in turn into the jig with the sensor head placed directly above it. The sensor head was lowered until contact was made with the sample and then increased in height up to 40~mm away in steps of 0.5~mm. At each step, a 0.5~s measurement was made. This measurement was repeated five times and an average line profile taken. Figure \ref{fig:SignalVsDistance} shows that the signal tends towards a base value with increasing distance. The magnetic field strength is inversely proportional to the distance from the source, $\mathrm{B \propto 1/r^n}$, where n is -1 for a current carrying wire and -3 for a dipole. A power law curve of the form $f(x) = a + bx^c$ was fitted to these measurements to check if these measurements followed the expected power law pattern and to see what was the value of the exponent. The fitted curve was also used to determine the maximum working distance of the magnetometer, for increasing concentrations. The fits showed an exponent value that started at -2 for higher concentrations and trended towards -1 at 10\%.

The iron-oxide suspension is super-paramagnetic because the size of the $\mathrm{\gamma-Fe_2O_3}$ iron-oxide nanoparticles is between 3.5 and 10~nm. Bulk $\mathrm{\gamma-Fe_2O_3}$ is ferrimagnetic due to ferric ions ($F_e^{3+}$),  with each ferric ion having a magnetic moment of 5~$\mu_B$ \cite{maghemite_magnetic_momoent}, where $\mu_B$ is the Bohr magneton. The bias field magnet on the magnetometer sensor head will induce a field in the sample as it gets closer, causing an increase in the detected magnetic field. This behavior is more complex than that of a simple dipole, which would have a $B \propto r^-{3}$ or $B \propto r^{-1}$ relationship for near and far field regimes respectively. The iron-oxide nanoparticles may also cluster together to form a more complex magnetic system. The magnetic field from the permanent bias magnet attached to the sensor has a field drop-off with distance also. The orientation of this field relative to the sample will also change how the field is induced in the sample.  The sample can be considered a finite volume of magnetic moments and during measurement the integrated contribution of all the individual moments across the sample is being detected. Due to the nano-size nature of the particles, they will be susceptible to Brownian motion while suspended in the solution which could cause the exponent value to deviate from $\mathrm{n = -3}$. The overall result is many individual dipoles, randomly distributed throughout a finite volume contributing non-uniformly to the total detected signal, causing a deviation from the -1 or -3 exponent value expected from a perfect dipole. 

To determine the point at which the signal from the sample is considered no longer detectable (the maximum working distance of the magnetometer) a threshold value of one standard deviation was used. For each concentration, the fitted curve was used to determine the point at which the curve went above the threshold value. The results are shown in Figure \ref{fig:MaxDistance}. As with the 2D scans, concentrations below 10\% are not shown as the SNR was too low. It is possible to conclude that the working distance decreases with concentration with the maximum working distance for 100\% concentration being 14.6 $\pm$ 2.2~mm and the 10\% concentration being the lowest concentration with a working distance above 0, at 5.8 $\pm$ 4.23~mm.

In conclusion, we have demonstrated an endoscopic fiber-coupled nitrogen-vacancy diamond DC magnetometer with a maximum sensor head diameter of 10~mm capable of detecting iron-oxide nanoparticles in MagTrace\texttrademark{} suspension. The lowest concentration detected was 2.8~mg/ml. The detection range of the lowest detectable concentration was 5.8 $\pm$ 4.2~mm. The maximum detection range of the probe was 14.6 $\pm$ 2.2~mm. Further improvements to the system could include a smaller sensor head design, taking advantage of the small size of the 0.5~$\mathrm{mm^3}$ diamond used for detection. This would help in making the magnetometer more applicable to scenarios where sensing is needed in confined spaces or to reduce physical trauma and obtrusiveness. Increased sensitivity would improve the detection range of iron oxide suspension as well as reduce the concentration that can be detected. Using more permanent magnets around the probe would increase the signal-to-noise by inducing a higher magnetic field in the sample.

\begin{acknowledgments}

Alex Newman's Ph.D studentship is funded by an EPSRC iCASE award to UKNNL (United Kingdom National Nuclear Laboratory). This work received funding from the National Nuclear Laboratory’s Science and Technology programme (Decontamination and Decommissioning Core Science). Stuart Graham's Ph.D studentship is funded by DSTL (the Defence Science and Technology Laboratory). 

This work is also supported by Innovate UK grant 10003146, EPSRC grant EP/V056778/1, EPSRC Impact Acceleration Account (IAA) award and the EPSRC Q-BIOMED Hub EP/Z533191/1. We thank the NHS for their help with this project, in particular Stuart Robertson and Joseph Hardwicke at the University Hospitals Coventry and Warwickshire (UHCW) for insightful discussions throughout the project. We also thank Douglas Offin from the National Nuclear Laboratory for useful conversations throughout this project.
\end{acknowledgments}

\appendix

\section{2D scans line profile fitting}

All 2D scans were summed together to determine where the average signal strength was across all scans. Due to the dipole shape of the signals, the 2D gradient of the image in the X direction was used to determine the centre of the dipole feature by finding the position where the gradient was the most negative. The X and Y coordinates of this point were used to draw a line profile vertically and horizontally through each of the individual 2D scans for each concentration. This then highlighted that, again due to the dipole nature of the signal, the vertical line profile passed through the centre of the dipole where there was no signal change. The decision was made then to only use the horizontal line profile. 

For each line profile, a linear baseline between the first and last data point was subtracted. This helped remove any any linear trend related to slow changes in signal over the scan that are caused by temperature changes or slowly varying magnetic fields. This also helped remove large static background field gradients across the line profile.


After the line profile was processed, a Gaussian derivative function of the form 

\begin{equation}
    f(x) = -\frac{A(x-\mu)}{\sigma^2}e^{-\frac{(x-\mu)^2}{2\sigma^2}} + c, 
\end{equation}

where A is the peak amplitude, $\mu$ is the peak position or zero crossing point, $\sigma$ is the standard deviation or width of the peak and $c$ is a signal offset.

The difference between the maximum and minimum values of the fit was taken as the the signal amplitude (\ref{signal_amp}). This value was divided by the standard deviation of the line profile data to find the SNR (\ref{SNR_eq}).

\begin{equation}
    f_{a} = f(x_{max}) - f(x_{min}),
    \label{signal_amp}
\end{equation}

\begin{equation}
    \mathrm{SNR} = \frac{f_{a}}{\mathrm{noise}}.
    \label{SNR_eq}
\end{equation}

The error on the SNR values was taken to be the sum of squares of the relative error of the signal amplitude and the signal noise,

\begin{equation}
    \sigma_{\mathrm{SNR}} = \mathrm{SNR} \times \sqrt{\left( \frac{\sigma_{f_{a}}}{f_{a}} \right)^2 + \left( \frac{\sigma_{\mathrm{noise}}}{\mathrm{noise}} \right)^2},
\end{equation}

where $\sigma_{f_{a}}$ is the error of the signal amplitude and $\sigma_\mathrm{noise}$ is the error of the signal standard deviation.

Firstly the error of the signal amplitude was found by using the square root of the sum of the squares of the error of the fit at the x position of the maximum and minimum values.

\begin{equation}
    \sigma_{f_{a}} = \sqrt{\sigma_{f(x_{max})}^2 + \sigma_{f(x_{min})}^2}
\end{equation}

The error of the fitting function was found using the square root of the sum of squares of the partial derivative of the fitting function with respect to each parameter

\begin{equation}
    \partial_A = \left( \frac{\partial f}{\partial A}\sigma_A \right)^2,
\end{equation}

\begin{equation}
    \partial_\mu = \left( \frac{\partial f}{\partial \mu}\sigma_\mu \right)^2,
\end{equation}

\begin{equation}
    \partial_\sigma = \left( \frac{\partial f}{\partial \sigma}\sigma_\sigma \right)^2,
\end{equation}

\begin{equation}
    \partial_c = \left( \frac{\partial f}{\partial c}\sigma_c \right)^2,
\end{equation}

\begin{equation}
        \sigma_f = \sqrt{\partial_A + \partial_\mu + \partial_\sigma + \partial_c}.
\end{equation}

The 95\% confidence intervals of the fitting parameters were used as the errors for each parameter. 

\section{Signal vs height line profile fitting}

Five height line profiles (line profiles moving in the z direction away from the sample) were taken for each concentration. Firstly each of the five height line profiles was baselined by subtracting the last data point, the signal furthest from the sample, from the line profile. The baselined data was then fitted with a power law function of the form



\begin{equation}
    f(x) = a + bx^c, 
\end{equation}

where $a$ is the baseline shift, $b$ is the peak amplitude and $c$ is the growth parameter. 

A value of the fitting parameter $a$ plus one standard deviation was used as a threshold value and the point at which the fitted curve went above this threshold value was used as the point at which the signal was deemed above the noise floor and therefore detected. This was called the maximum working distance. The standard deviation value was found by measuring the background signal (i.e. no sample in place) over the same height line profile and taking the standard deviation. 

This was repeated for each of the five profiles taken. The average maximum working distance was then used as well as the standard deviations of these distances as the error.

\section{Larger sensor head design and changes in ODMR spectra}

\begin{figure}[t]
    \centering
    \includegraphics[width=\linewidth]{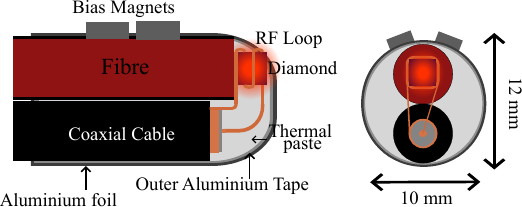}
    
    \caption{A previous design that was larger but still endoscopic utilized two bias magnets, aluminum foil and tape to help with reflecting fluorescence and making the sensor head light tight.}
    \label{fig:old_sensor_head_design}
\end{figure}

Previous designs to the one shown in the main paper used two bias magnets placed back from the diamond but not in line with each other. The magnets were rotated around the diameter of the fiber in opposite directions (Fig. \ref{fig:old_sensor_head_design}). This was again to achieve a [100] bias field alignment with the diamond. The diamond, although still glued directly to the front of the fiber, was in a different position than where it was for the work in the main paper. 

\begin{figure}[b]
    \centering
    \begin{subfigure}{0.3\textwidth}
        \phantomsubcaption
        \label{fig:old_100p_2d_scan}
        \includegraphics[width = \textwidth]{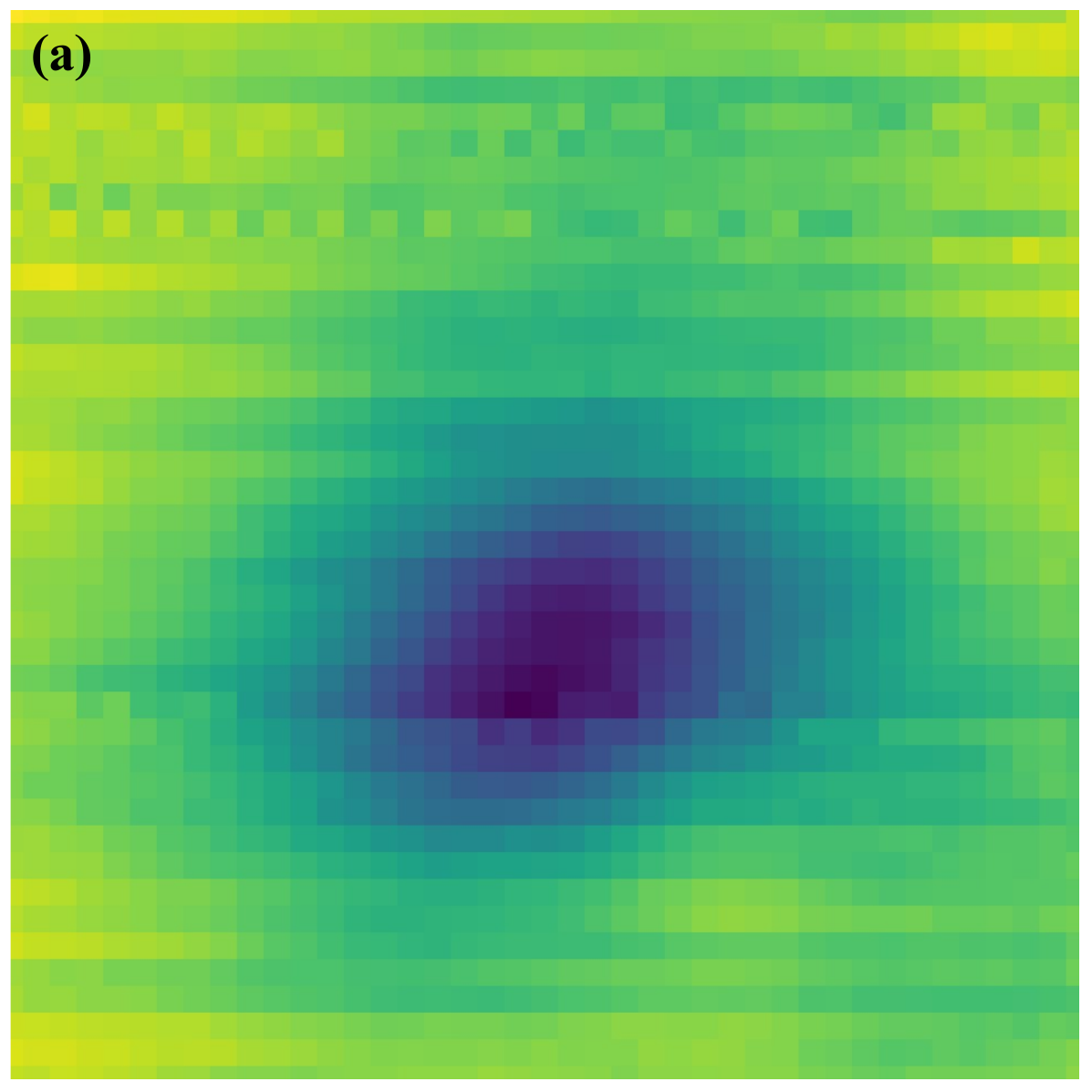}
    \end{subfigure}
    \begin{subfigure}{0.45\textwidth}
        \phantomsubcaption
        \label{fig:old_100p_line_profile}
        \centering
        \includegraphics[width = \textwidth]{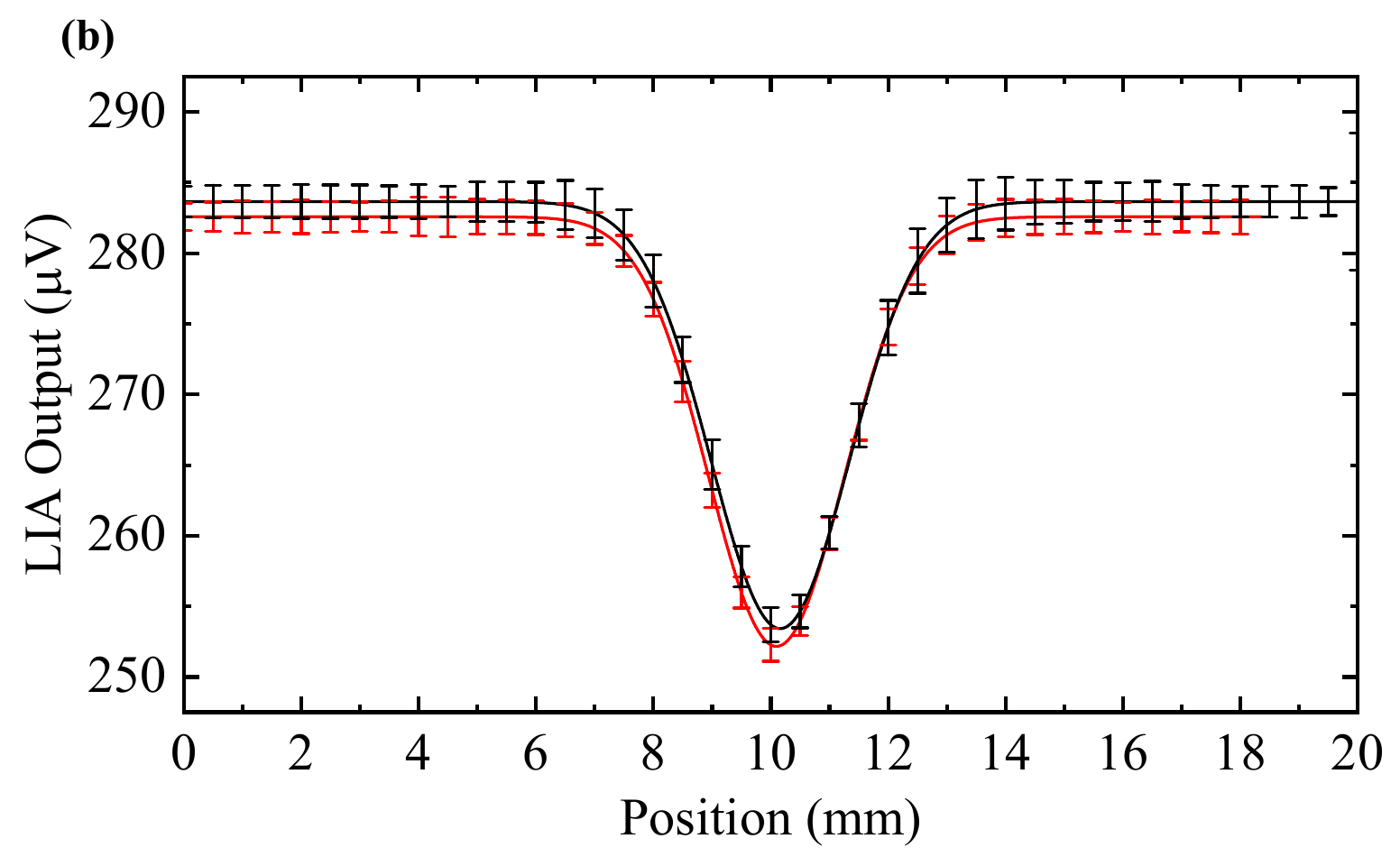}
    \end{subfigure}
    \caption{A change in the magnet configuration results in a change in the measured field pattern while scanning over samples. (a) A 20 x 20 mm scan of a 100\% concentration sample using the old probe configuration. (b) The Gaussian fits for the horizontal (black) and vertical (red) line profiles over the signal. }
    \label{fig:old_scan_data}
\end{figure}

\begin{figure}[t]
    \centering
    \begin{subfigure}{0.5\textwidth}
        \phantomsubcaption
        \label{fig:simODMRModDepthcomp}
        \includegraphics[width = \textwidth]{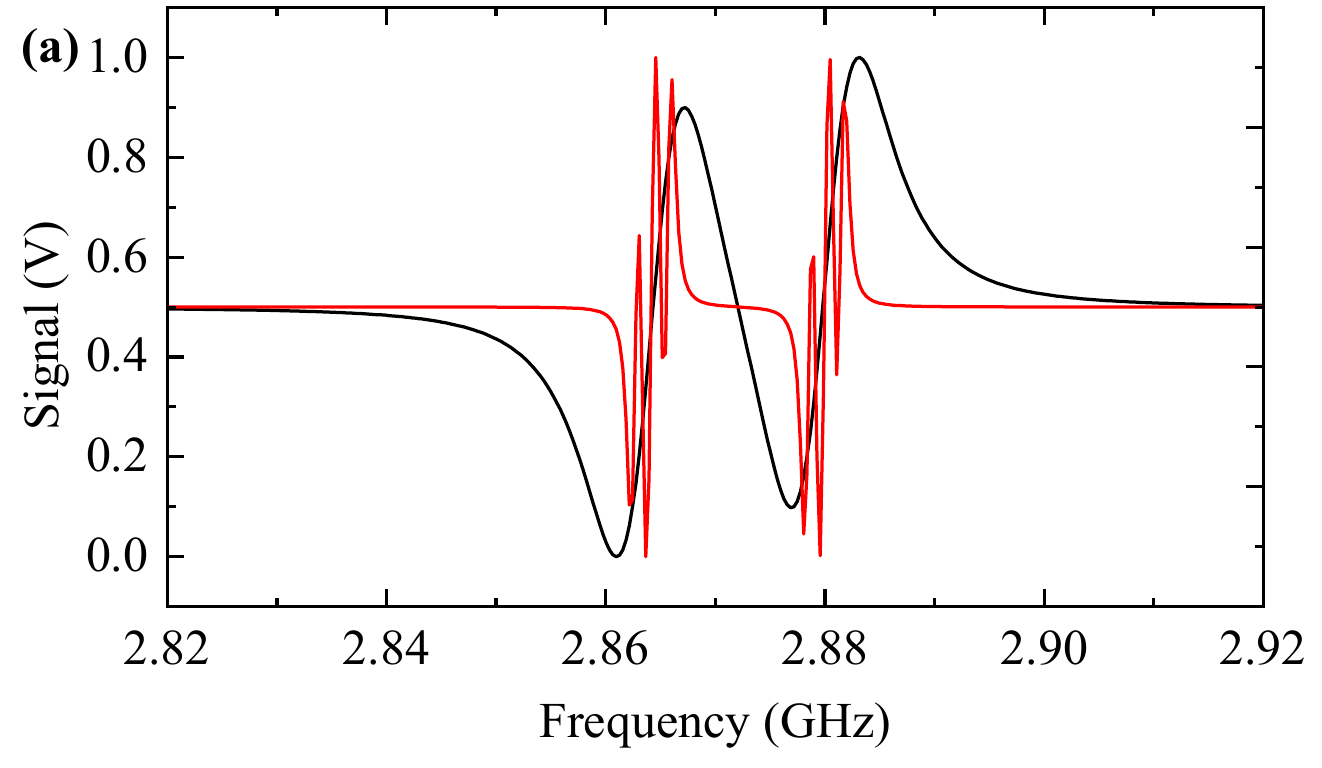}
    \end{subfigure}
    \begin{subfigure}{0.5\textwidth}
        \phantomsubcaption
        \label{fig:simODMRModDepthcomp_SigRes}
        \centering
        \includegraphics[width = \textwidth]{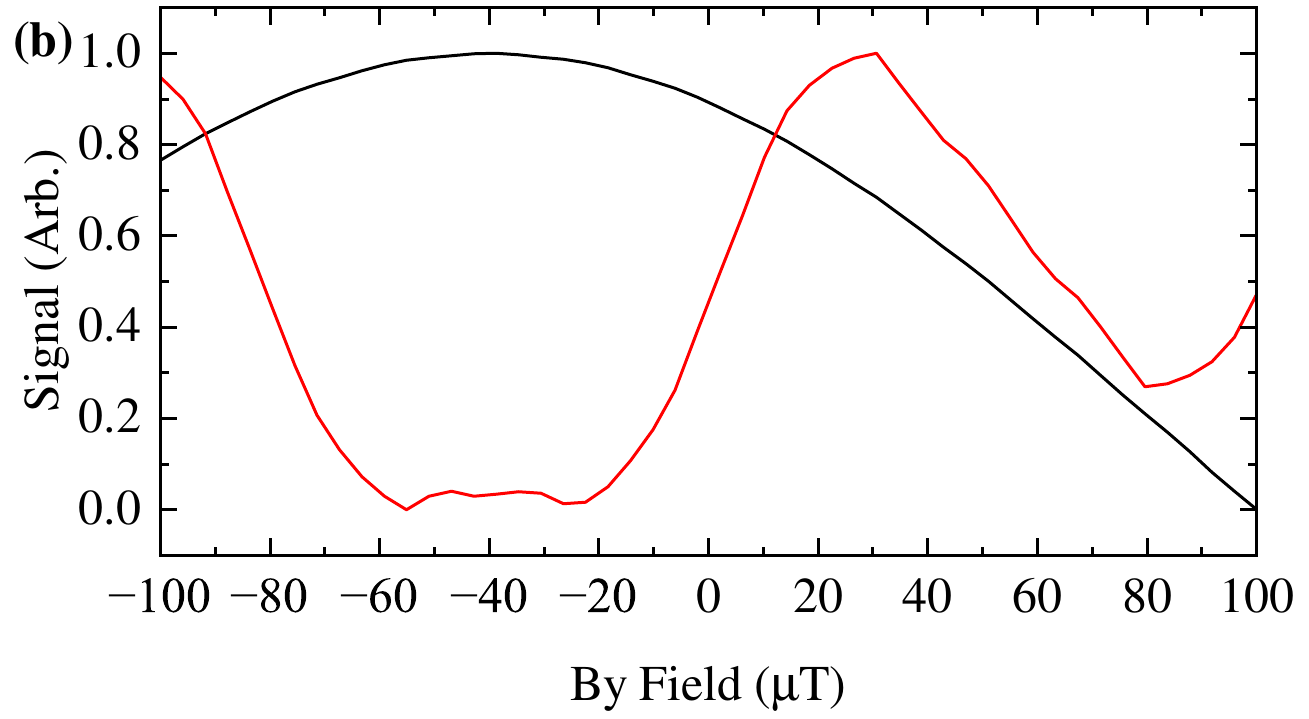}
    \end{subfigure}
    \caption{By simulating the ODMR spectra of an imperfect [100] alignment with 500~kHz and 5~MHz broadened peaks, the overall shape and therefore the signal response due to external fields changes. (a) The simulated ODMR spectrum for 500~kHz (red) and 5~MHz (black) peak broadening (b) The signal response at the same point in both ODMR spectrum, as an applied $\mathrm{B_y}$ field is swept between -100 $\mu T$ and 100 $\mu T$.}
    \label{fig:simODMR}
\end{figure}

This configuration of magnet position relative to the probe head, diamond and the samples caused a different response in fluorescence where a Gaussian dip was seen rather than a derivative shape (Fig. \ref{fig:old_scan_data}). The pair of magnets were positioned so that both the north poles were in the same direction, which resulted in an increase in overall field amplitude around the sensor head. This increase in field around the sensor tip would have caused an induced field on the sample that was larger than with the current single magnet setup, producing a larger response signal to detect. The extra magnet was removed however to reduce the overall diameter of the probe. Although the induced field would have been larger, the measurement contrast and maximum working distance calculated from this data set was lower than that of this current configuration. This was due to the broader ODMR spectrum obtained from this configuration. Although the bias field alignment with the two magnets was such that a [100] alignment was seen, a much broader frequency modulation of 5~MHz amplitude was used, compared with 0.5~MHz for the current sensor head design. This broader ODMR spectrum caused the fluorescence response to be both weaker and non-linear due to the convolution of the now broadened individual peaks creating a single broad peak (Fig. \ref{fig:simODMRModDepthcomp}).

    

The ODMR used in Fig.\ref{fig:ODMR}, where a single peak can be selected at a particular frequency that corresponds to one of four possible sensitive axes in the diamond, responds proportionally to the field strength along the corresponding NV symmetry axis as described by the Zeeman effect. However with the single broad peak being made of an overlap of four separate broad peaks, signal response is not simply from the increase or decrease in resonance frequency but is now due to the four separate peaks moving up and down the resonance frequency range in different directions relative to each other. This then means the signal response measured is due to the overall broadness of the single peak changing as the convolution of these four separate peaks changes. A simulation that solves the NV centre ground state Hamiltonian to calculate the resonance frequencies for an applied bias field and test field was used to generate ODMR spectra. The ODMR simulation was used to generate spectra for both a 0.5~MHz and a 5~MHz line width, to simulate broadening from an increased modulation amplitude. An initial [100] bias field with a small off-axis field was applied, to simulate the imperfect position of the bias magnet on the sensor head (Fig. \ref{fig:simODMRModDepthcomp}). The ODMR with the 5~MHz line width was first generated and then a measurement point at the zero-crossing resonance frequency of the first peak was chosen. An applied test field along each of the three magnetic field axes of -100 $\mu T$ to 100 $\mu T$ was applied and the change in signal at the zero-crossing point was measured (Fig. \ref{fig:simODMRModDepthcomp_SigRes}). Sweeping the $\mathrm{B_y}$ component, for example, is dramatically different for the broad single peak compared to the many unbroadened peaks.

These changes in signal response due to the difference in modulation amplitude are most likely the reason for the difference between the two sets of data, becoming the most dominant effect over the increase in signal strength due to the addition of an extra magnet.

\section{Sensitivity}

\begin{figure}[t]
    \centering
    \includegraphics[width=\linewidth]{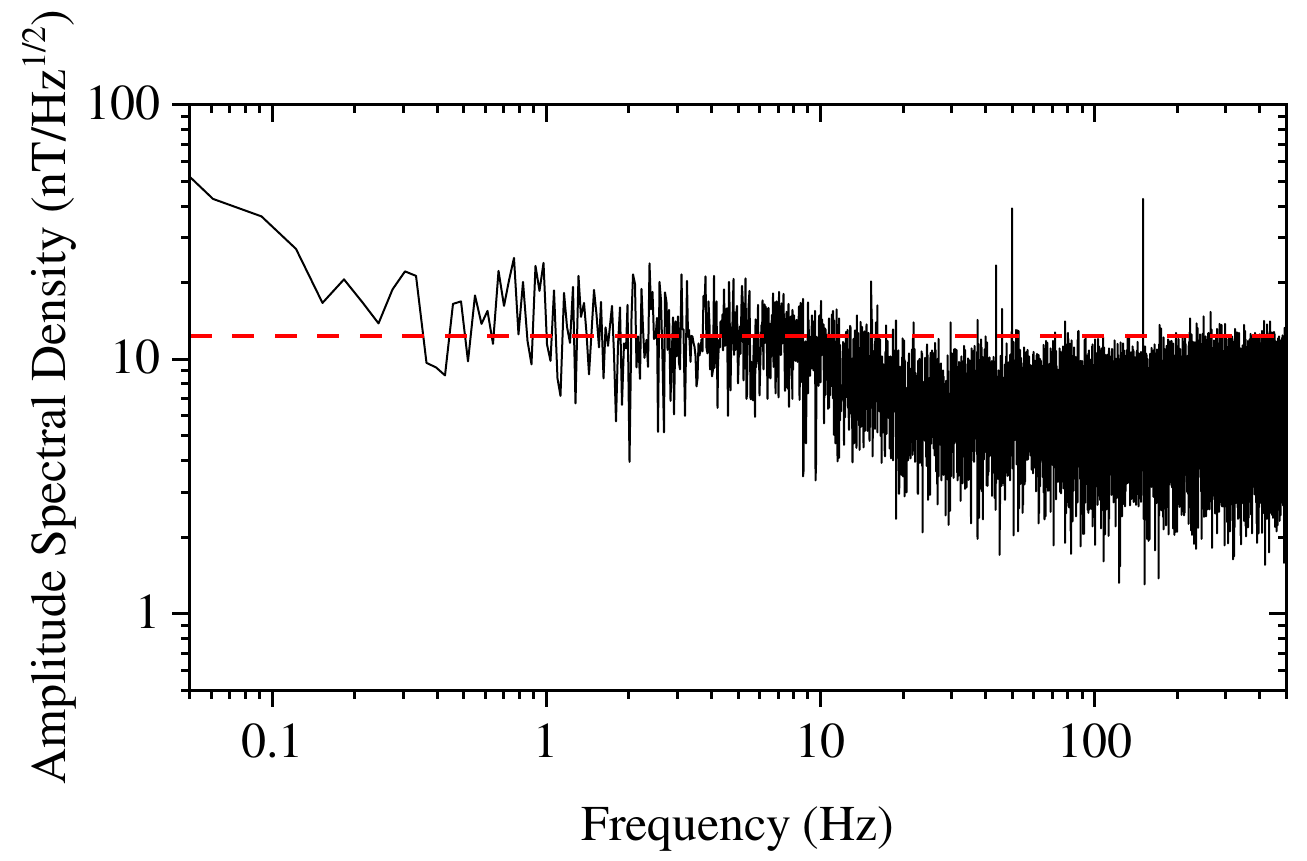}
    \caption{The average amplitude spectral density from the fast-Fourier-transforms (FFT) of three time traces that each lasted 16 minutes and 20 seconds. The red dashed line shows the average value between 0.5 and 10~Hz of $\mathrm{12.3\pm4.1~nT/Hz^{1/2}}$.}
    \label{fig:probe_sens}
\end{figure}

To determine the sensitivity of the magnetometer, three fast-Fourier-transforms (FFTs) were measured while the microwave frequency was set to the zero-crossing point of the central peak of the left-hand side group of peaks (the $m_s = -1$ transition peaks) in the ODMR spectrum. The FFTs were taken using a PicoScope 5442D at a sampling rate of 1~kHz. The measured signal was amplified by 500 times by the lock-in amplifier before going into the PicoScope. The average of the three FFTs was used to determine the average noise value between 0.5 and 10~Hz as shown in Fig.\ref{fig:probe_sens}. 10~Hz was the upper limit as a low-pass filter with a 3~dB point of 10~Hz was used to reduce noise. Each FFT was taken with a rectangular window applied. This average value in units of dBu was converted to volts by $V = V_{ref} \times 10^{\mathrm{dBu}/20}$ with the reference voltage $V_{ref} = 0.775$~V. This unit of voltage was then converted to units of nT using the gradient of the chosen ODMR peak and the gyromagnetic ratio for the NV centre. The gradient of the peak was measured using a linear fit to be approximately 0.22~V/MHz and the gyromagnetic ratio of the NV centre is know to be 28 GHz/T. These two units then give a calibration constant of $6.16\times 10^{-6}$ V/nT. Using this calibration constant, the average FFT value was found to be $\mathrm{12.3 \pm 4.1~nT/Hz^{1/2}}$ between 0.5 and 10~Hz. This was taken to be the unshielded sensitivity of the magnetometer.


%

\end{document}